\documentstyle[12pt,psfig,cite]{article}
\begin{document}
\renewcommand{\thefootnote}{\fnsymbol{footnote}}
\begin{titlepage}
\vspace*{1cm}
\Large
\begin{center}
Scaling Behavior in 4D Simplicial Quantum Gravity
\end{center}
\vspace{0.5cm}
\begin{center}
\large
H.S.Egawa, \footnote[2]{E-mail address: egawah@theory.kek.jp} \\
\normalsize
Department of Physics, Tokai University, \\
Hiratsuka, Kanagawa 259-12 \\
\vspace{5mm}
\large
T.Hotta, \footnote[3]{E-mail address: hotta@danjuro.phys.s.u-tokyo.ac.jp}
T.Izubuchi, \footnote[4]{E-mail address: izubuchi@danjuro.phys.s.u-tokyo.ac.jp} 
\\
\normalsize
Department of Physics, University of Tokyo, \\
Bunkyo-ku, Tokyo 113 \\
\vspace{5mm}
\large
N.Tsuda \footnote[5]{E-mail address: ntsuda@theory.kek.jp} \\
\normalsize
National Laboratory for High Energy Physics (KEK) \\
Tsukuba, Ibaraki 305 , Japan \\
\vspace{5mm}
\large
and \\
\vspace{5mm}
T.Yukawa \footnote[6]{E-mail address: yukawa@theory.kek.jp} \\
\normalsize
Coordination Center for Research and Education, \\
The Graduate University for Advanced Studies, \\
Hayama-cho, Miura-gun, Kanagawa 240-01, Japan \\
and \\
National Laboratory for High Energy Physics (KEK) \\
Tsukuba, Ibaraki 305, Japan
\end{center}

\vspace{0.5cm}

\begin{abstract}
Scaling relations in four-dimensional simplicial quantum gravity are 
proposed using the concept of the geodesic distance.
Based on the analogy of a loop length distribution in the two-dimensional 
case, the scaling relations of the boundary volume distribution in four 
dimensions are discussed in three regions: the strong-coupling phase, the 
critical point and the weak-coupling phase. 
In each phase a different scaling behavior is found.
\end{abstract}

\end{titlepage}

\section{Introduction}
Two-dimensional quantum gravity has generally been regarded as a toy-model 
theory of four-dimensional quantum gravity. 
According to the Liouville theory,\cite{KPZ,DK} two-dimensional quantum 
gravity can be quantized with a central charge $c < 1$.  
The method of dynamical triangulation\cite{DT} has generally been known 
as a correct discretized model corresponding to the matrix model, and 
has the same critical index as Liouville field theory in the continuum 
limit. 
In two dimensions many attempts, for example the so-called minbu analysis,
\cite{MINBU1,MINBU2} the loop length distributions,\cite{TY,KKMW,IK} 
the fractal dimensions and an analysis of the complex structures\cite{CPS} 
reveal the many important properties of random surfaces.
Especially, we take notice of the concept of the loop length 
distribution.
Fortunately, the loop length distribution function has been 
calculated analytically for the case of two-dimensional pure gravity.
\cite{KKMW,IK}
Our basic strategy in four dimensions is that higher-dimensional quantum 
gravity can be represented as a superposion of low-dimensional quantum 
gravity.\cite{SAVVIDY2}
Indeed, a two-dimensional random closed surface can be reduced to a 
summation over the loops along a certain gauge slice regarded as the 
geodesic distance corresponding to time in our case. 
One of the aims of this article is to investigate four-dimensional 
Euclidean spacetime structures using the geodesic distance (precisely, 
the extended concept of the loop length distribution in two-dimensional 
simplicial quantum gravity).
Based on an analogy of the loop length distribution in the two-dimensional 
case, the scaling relations in the four-dimensional case are discussed for 
three regions $i.e.$ the strong-coupling phase, the critical point and 
the weak-coupling phase. 
In the three-dimensional case the scaling relations of the boundary-area 
distributions have already been argued using the analogy of the loop length 
distribution, resulting in the establishment of scaling properties.\cite{HTY}

In order to discuss the scaling relations of the boundary volume, we assume 
that the scaling variable ($x$) has the form $V/D^{\alpha}$, where $V$, $D$ 
and $\alpha$ denote each boundary volume, the geodesic distance and the 
scaling parameter, respectively. 

The rest of the article is organized as follows: section 2 briefly reviews 
the definition of the model.
Section 3 considers the new scaling relations of the boundary volume 
obtained in four dimensions, and the following subsections look at these in 
detail for the three regions.
Finally, section 4 contains a summary of the results and discussions. 

\section{The model}
It is still not known how to give a constructive definition of 
four-dimensional quantum gravity.
We naturally must mention the Euclidean Einstein-Hilbert action, as follows: 
\begin{equation} 
S_{EH} = \int d^4x \sqrt{g} \left(\Lambda - \frac{1}{G}R \right), 
\end{equation}
where $\Lambda$ is the cosmological constant and $G$ is Newton's constant 
of gravity.
We use the lattice action of the four-dimensional model with the 
topology $S^4$, corresponding to the above action, as follows: 
\begin{eqnarray} 
S(\kappa_{2},\kappa_{4}) & = & - \kappa_2 N_2 + \kappa_4 N_4  
          \nonumber\\    
  & = & - \frac{2\pi}{G}N_2 
        + \left( \Lambda' + \frac{10}{G} cos^{-1}(\frac{1}{4})
          \right)N_4, 
\end{eqnarray}
where $N_i$ denotes the total number of $i$-simplexes and 
$\Lambda' = c\Lambda$; $c$ is the unit volume and 
$cos^{-1}(\frac{1}{4})$ is the angle between two tetrahedra.
The coupling $\kappa_2$ is proportional to the inverse bare 
Newton's constant, and the coupling $\kappa_4$ corresponds 
to a lattice cosmological constant. 

For the dynamical triangulation model of four-dimensional 
quantum gravity we consider a partition function of the form 
\begin{equation} 
Z(\kappa_2, \kappa_4) 
= \sum_{T(S^4)} e^{-S(\kappa_2, \kappa_4)}.
\label{eq:PF} 
\end{equation}
We sum over all simplicial triangulations, $T(S^4)$, on a four-dimensional 
sphere.
Here, we fix the topology as $S^4$. 
In practice, we must add a small correction term,\cite{Agis_Migd} 
$\Delta S $, to the lattice action in order to suppress volume 
fluctuations. 
The correction term is denoted by  
\begin{equation} 
\Delta S = \delta \kappa_4 (N_4 - N_4 ^{(target)})^2,
\end{equation}
where $N_4 ^{(target)}$ is the target value of four-simplexes, 
and $\delta = 0.0005$ is used.
%
\section{Numerical simulation and results}
\begin{figure}
\centerline{\psfig{file=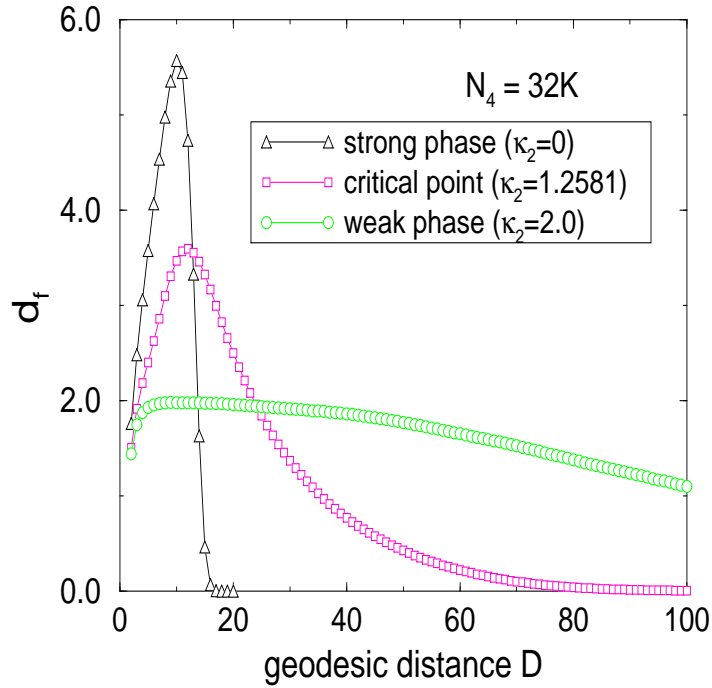,height=10cm,width=10cm}}
\caption
{
Fractal dimensions ($d_{f}$) versus the geodesic distance ploted for the 
strong phase, critical point and weak phase with $N_{4}=32K$.
}
\label{Fig:FD}
\end{figure}
%
\begin{figure}
\centerline{\psfig{file=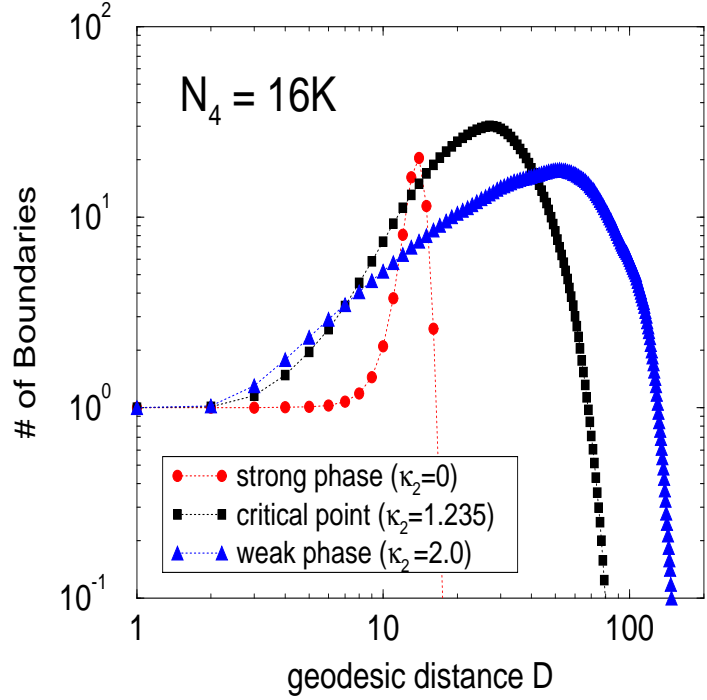,height=10cm,width=10cm}}
\caption
{
Number of boundaries versus the geodesic distance for the strong phase, 
critical point and weak phase with double-log scales.
}
\label{Fig:NB}
\end{figure}
%
One of the interesting observables in this model is the fractal 
dimension ($d_{f}$) at large scale.
A common way to define the fractal dimension is based on studying the 
behavior of the volume within a geodesic distance $D$, ($V(D)$).
Usually, $d_{f}$ is calculated as the following quantity: 
\begin{equation}
d_{f} \equiv \frac{d\ln{V(D)}}{d\ln{D}},
\label{eq:fracta_dimension}
\end{equation}
which would be a constant if $V(D)$ goes like a power of $D$.
Fig.\ref{Fig:FD} denotes that the fractal dimensions are plotted versus 
the geodesic distances with $N_{4}=32K$.
We have to inquire, to some extent, into the internal geometrical 
view points of the four-dimensional dynamically triangulated 
manifolds.

We define $N_{b}(D)$ as the number of boundaries at a geodesic 
distance $D$ from a reference four-simplex in the dynamically 
triangulated manifolds.
Fig.\ref{Fig:NB} shows the distributions of $N_{b}(D)$ for the typical 
three coupling strengths with $N_{4}=16K$.
In the strong-coupling limit ($\kappa_{2}=0$, this corresponds to 
$G \rightarrow \infty$), the only boundary that is identified as the 
mother universe exists within almost all of the geodesic distances; also the 
creation of the branches is highly suppressed, which means that the mother 
universe is a dominant structure.
The suppression of branches shows one of the characteristic properties 
of a ``crumpled manifold'', which is similar to the case observed in the 
two-dimensional manifold. 
On the other hand, in the weak-coupling phase (for example, we chose 
$\kappa_{2}=2.0$, (see Fig.\ref{Fig:NB})), we see the growth of the 
branches (``elongated manifold'') until $D \sim 60$ for $N_{4}=16K$.
From Fig.$1,2$ there is no doubt that the dynamically triangulated 
manifold becomes crumple in the strong-coupling phase and a branched 
polymer in the weak-coupling phase.\cite{BS_4DC,Scal_4DQG}
%


We now devote a little more space to explaining the loop length 
distributions in two-dimensional dynamically triangulated surfaces.
We suppose a disk which is covered within a geodesic distance of 
$D$ from the arbitrary triangle.
Because of the branching of the surface, a disk is not always 
simply-connected, and there usually appear many boundaries consisting of 
$S^{1}$ loops in this disk.
The loop length distribution function ($\rho(L,D)$) is measured by 
counting the number of boundary loops with the loop length ($L$) within 
a geodesic distance ($D$).
In ref.7) the loop length distribution function ($\rho(L,D)$) has been 
calculated as a function of the scaling variable, $x = L/D^{2}$, 
in the continuum limit for the case of pure gravity in two dimensions, 
\begin{equation}
\rho(x=\frac{L}{D^{2}},D) = \frac{3}{7\sqrt{\pi}} \frac{1}{D^{2}} 
\left( x^{-5/2} + \frac{1}{2}x^{-3/2} + \frac{14}{3} x^{1/2} \right) 
e^{-x}.
\label{eq:LLD}
\end{equation}
This distribution function consists of two different types of 
distributions.
The first two terms of eq.(\ref{eq:LLD}), 
$(x^{-5/2} + \frac{1}{2}x^{-3/2})e^{-x}$, represent the so-called baby 
loops.
The baby loops originate from the quantum fluctuation of the surface.

The last term of eq.(\ref{eq:LLD}), $(\frac{14}{3} x^{1/2}) e^{-x}$, 
represents the mother loop.
The precise definition of the mother loop is the boundary of the largest 
uncovered regions for each geodesic distance.
One of the notable features of the loop length distribution is that the 
mother loop and the baby loops distribute with the same scaling 
variable, $x=L/D^{2}$.
There are also other things to note.
The distribution function of the baby loops gives the divergences in the 
calculations of the number of boundaries at distance $D$, 
$M^{0}(D) = \int_{0}^{\infty} d L \; \rho(L,D)$, and the total length 
of boundaries at distance $D$, 
$M^{1}(D) = \int_{0}^{\infty} d L \; L \; \rho(L,D)$.
These divergences indicate that $M^{0}(D)$ and $M^{1}(D)$ depend on 
the lattice cut-off in the model, which offers a key to understanding 
the universality of the distributions.

\begin{figure}
\centerline{\psfig{file=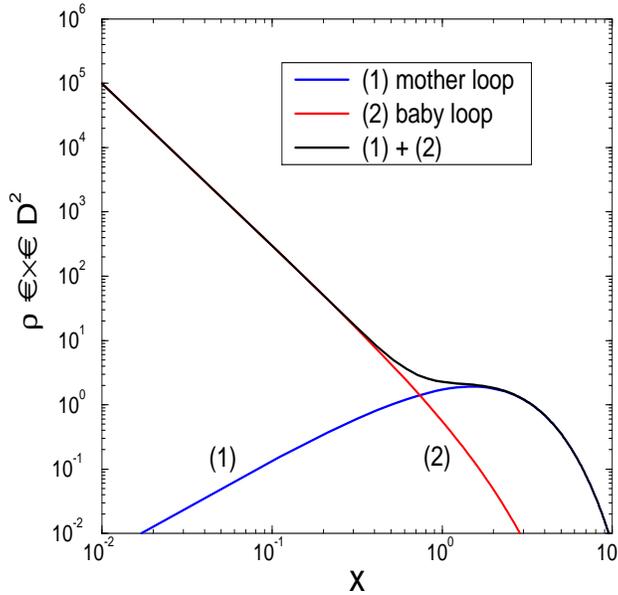,height=9cm,width=9cm}}
\caption
{
Loop length distribution function ($\rho(x,D)$), consisting of 
the mother loop and baby loops distribution function.
(1)mother loop distribution function: $\frac{14}{3} x^{1/2} e^{-x}$ 
(thick line) and 
(2)baby loop distribution function: $(x^{-5/2} + \frac{1}{2} x^{-3/2}) 
e^{-x}$ (dotted line).
Both are plotted with double-log scales.
}
\label{Fig:LLD_Theory}
\end{figure}

For future convenience, we show the mother and baby loop distributions in 
Fig.\ref{Fig:LLD_Theory}.
Similar distributions also appear in four dimensions!
%
%
Numerically, these distribution functions were investigated in ref.6); 
excellent agreement between the numerical results and the analytical 
approaches has been obtained.
In the sense that two distributions of the sections (or boundaries) of 
the surface at different distances are exactly the same as each other 
after a proper rescaling of the loop length, we can safely state that 
the dynamically triangulated surfaces become fractal.
That is to say, $\rho$ satisfies relation (\ref{eq:scaling}) under 
rescaling, $D \rightarrow D'=\sqrt{\lambda}D$ and 
$L \rightarrow L'=\lambda L$, 
\begin{equation}
\rho(x,D) \equiv \rho(L,D) = \lambda^{-1} \rho(L', D').
\label{eq:scaling} 
\end{equation}

We must now return to the four-dimensional case.
The sections (or boundaries) of a four-dimensional dynamically triangulated 
manifold are closed three-manifolds.
The boundary volume distribution ($\rho(x,D)$) is defined based on an analogy 
of the loop length distribution in two dimensions, and assuming that the 
scaling variable($x$) has the form $V/D^{\alpha}$, where $D$ denotes 
the geodesic distance in the four-dimensional dynamically triangulated 
manifold. 
%

\subsection{The strong-coupling phase}
\begin{figure}
\centerline{\psfig{file=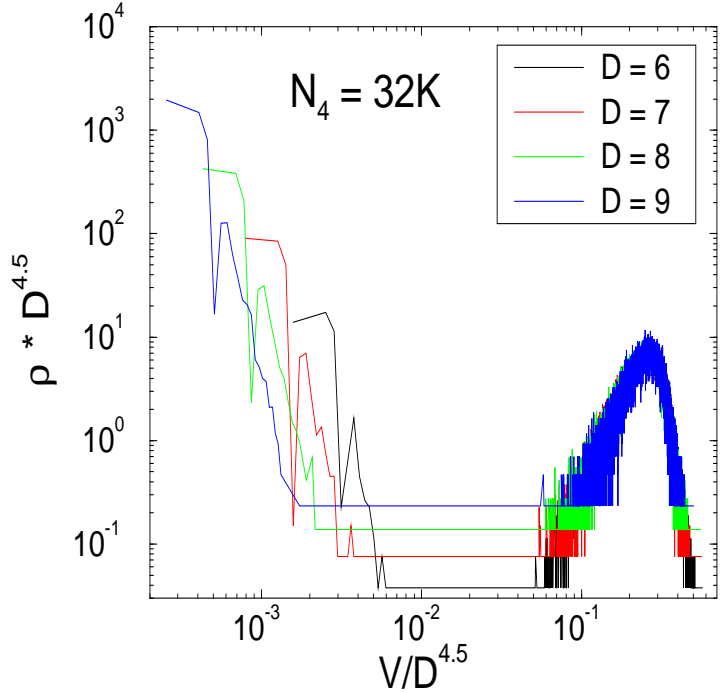,height=10cm,width=10cm}}
\caption
{
Boundary volume distribution in the strong-coupling limit with 
$N_{4} =32K$ with double-log scales.
}
\label{Fig:BVD_Strong}
\end{figure}

\begin{figure}
\centerline{\psfig{file=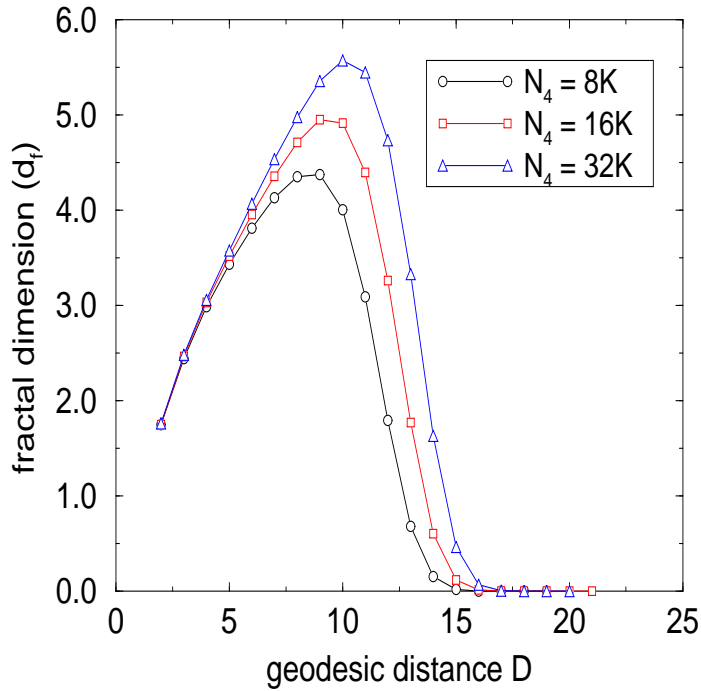,height=10cm,width=10cm}}
\caption
{
Fractal dimensions plotted versus the geodesic distance in the strong-coupling 
limit, $\kappa_{2}=0$ with $N_{4}=8K, 16K, 32K$.
}
\label{Fig:Strong_fractal_dim}
\end{figure}

Fig.\ref{Fig:BVD_Strong} shows the boundary volume distributions, 
$\rho(x) \times D^{4.5}$, with $x = V/D^{4.5}$ as the scaling variable in the 
strong-coupling limit, $\kappa_{2} = 0$, while the fractal dimension($d_{f}$) 
reaches $5.5$ with $N_{4}=32K$.
The boundaries of the mother universe show a good scaling property with 
$x = V/D^{d_{f}-1}$ as the scaling variable, and have a Gaussian distribution.
We can recognize that in the strong-coupling phase the scaling parameter 
($\alpha$) of the mother universe satisfies the relation 
$d_{f} = \alpha + 1$, and that the manifold resembles a $d_{f}$-sphere
($S^{d_{f}}$),\cite{BS_4DC} where $d_{f}$ denotes the fractal dimension, 
which increases with the volume\cite{Scal_4DQG}(see 
Fig.\ref{Fig:Strong_fractal_dim}).

On the contrary, we have no evidence concerning the scaling properties of 
the volume of the baby boundaries.
To be precise, it is impossible to scale the baby volumes in the same manner 
as the mother volume.
We find a rapid increase in the distribution of the baby volumes when 
$x \rightarrow 0$.
This rapid increase indicates the existence of the ``Planck scale'' in the 
model.
%

%
%

\subsection{The critical point}
\begin{figure}
\centerline{\psfig{file=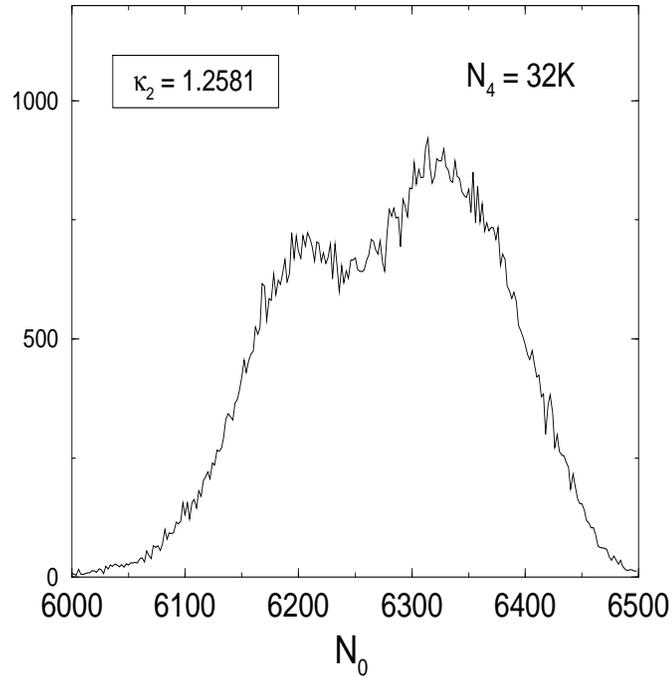,height=10cm,width=10cm}}
\caption
{
Histogram of $V_{0}$ at the critical point $\kappa_{2} = 1.2581$ with 
$N_{4} = 32K$.
}
\label{Fig:Double_Peak}
\end{figure}

\begin{figure}
\centerline{\psfig{file=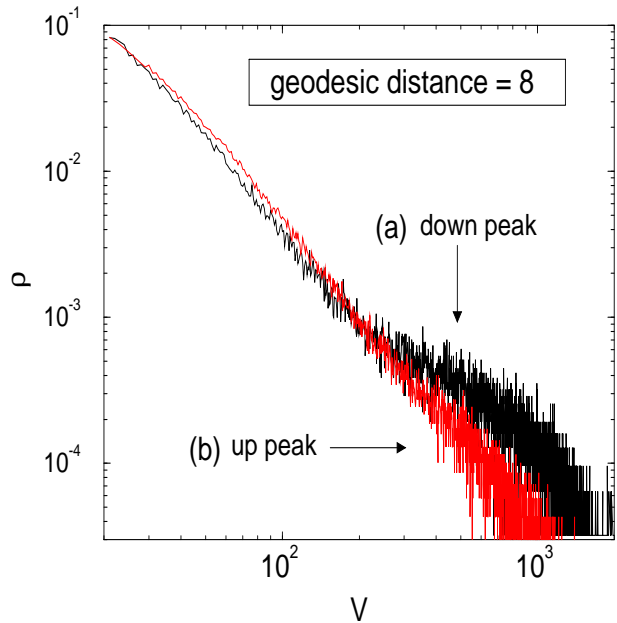,height=9cm,width=9cm}}
\caption
{
Boundary volume distributions of the two peaks with double-log scales.
$(a)$ denotes the boundary volume distributions of the ``down peak'' 
and $(b)$ denotes the boundary volume distributions of the ``up peak''.
In the case of $(b)$, the mother volume distribution almost disappears.  
}
\label{Fig:BVD_32K_UP_DN}
\end{figure}

We have long believed that the phase transition in four dimensions is 
continuous. 
According to recent reports,\cite{BBKP_4DQG} the phase transition of 
four-dimensional simplicial quantum gravity seems to be first order.
Indeed, we have observed the double-peak structure of a histogram of 
$N_{0}$ in our measurements at a volume of $N_4 = 32K$ and 
$\kappa_{2} = 1.2581$ (see Fig.\ref{Fig:Double_Peak}), which is a sign 
of a first-order transition, contrary to what is generally believed.
We must draw attention to the double-peak histogram structure on the 
critical point. 
We thus measure the boundary volume distribution on both peaks, 
and obtain the more clear signal of the mother universe on the peak 
which is close to the strong-coupling phase.
Fig.\ref{Fig:BVD_32K_UP_DN} shows the boundary volume distribution on 
two 
peaks.
We call the peak close to the strong-coupling phase the ``down peak'', 
and the other peak close to the weak-coupling phase the ``up peak''.

%
%
\begin{figure}
\centerline{\psfig{file=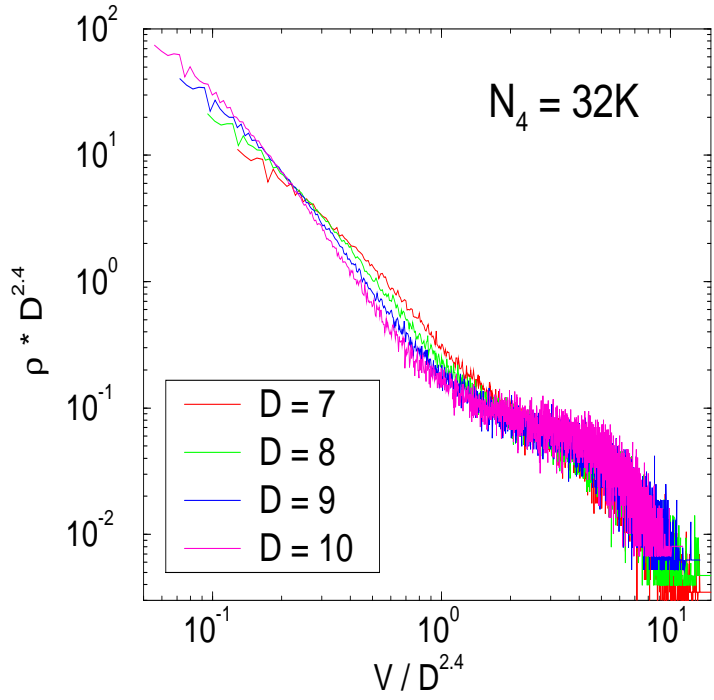,height=10cm,width=10cm}}
\caption
{
Boundary volume distributions with scaling variable $x=\frac{V}{D^{2.4}}$ 
at the critical point $\kappa_{2} = 1.2581$ with double-log scales.
}
\label{Fig:BVD_Critical}
\end{figure}
%
\begin{figure}
\vspace{-0.5cm}
\centerline{\psfig{file=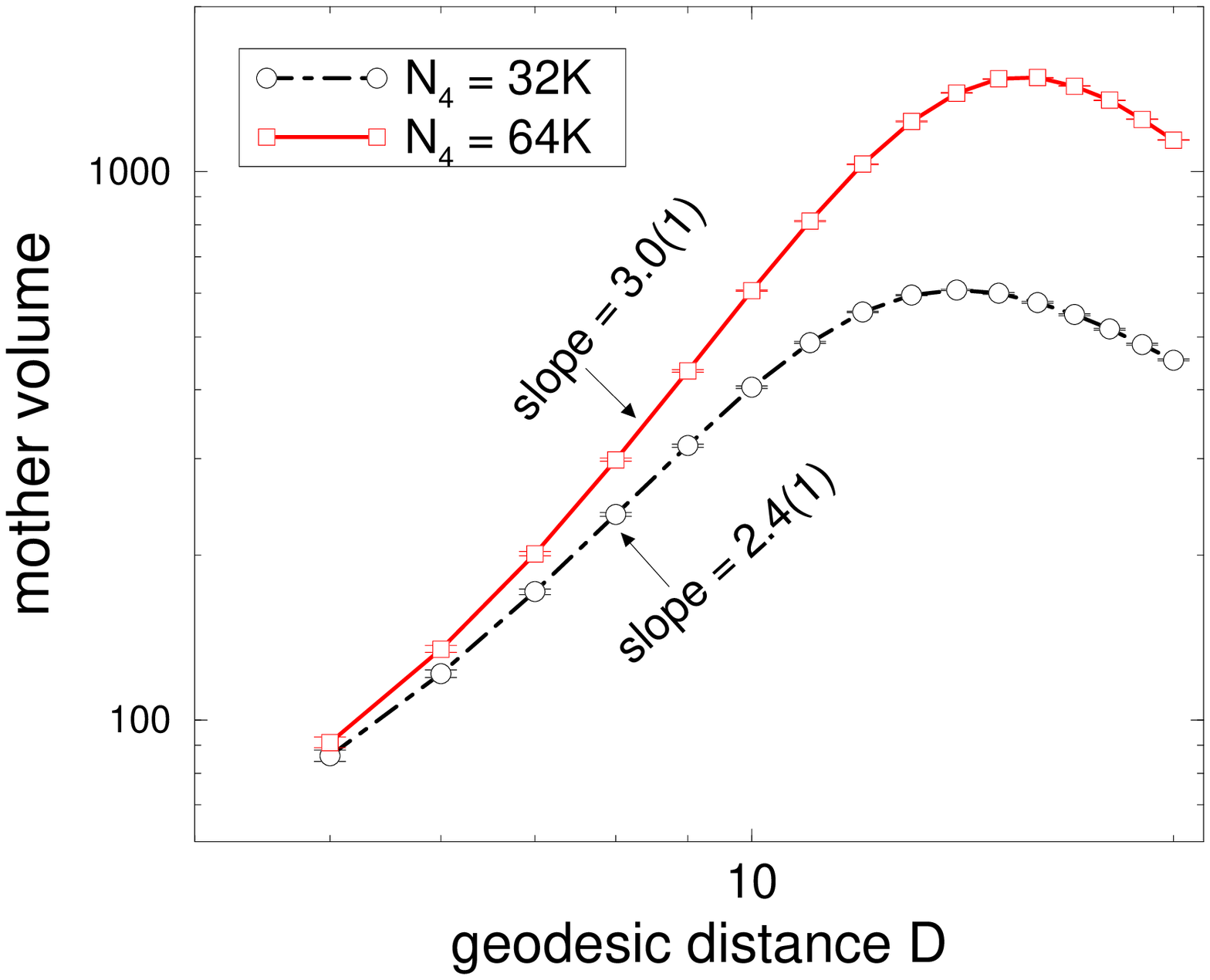,height=9cm,width=9cm}}
\caption
{
Volume of the mother boundary versus the geodesic distance with double-log 
scales at the critical point $\kappa_{2}=1.2581$ and $1.2770$ with 
$N_{4}=32K$ and $64K$, respectively.
For the case of $N_{4}=32K$ we estimate the scaling parameter 
$\alpha = 2.4(1)$ for the range $7 \leq D \leq 10$.
For the case of $N_{4}=64K$ we also estimate the scaling parameter, 
$\alpha = 3.0(1)$, for the range $7 \leq D \leq 12$.
}
\label{Fig:BVD_Mother_Critical}
\vspace{-0.5cm}
\end{figure}

The distributions of the boundary volume obtained from the ``up peak'' are 
similar to the distributions obtained in the weak-coupling phase (see next 
subsection).
That is to say, the structure of the mother volume disappears 
(see Fig.\ref{Fig:LLD_Theory},\ref{Fig:BVD_32K_UP_DN}).
There is fairly good general agreement that the dynamically triangulated 
manifold behaves like a branched polymer in the weak-coupling phase. 
We may therefore say that the boundary volume distributions of the 
``up peak'' are non-universal.

The data obtained in the ``down peak'' are shown in 
Fig.\ref{Fig:BVD_Critical} for various geodesic distances(D).
We observe two different scaling properties: one is for the mother 
boundary, and the other is for the baby boundaries. 
What we have learned in two-dimensional quantum gravity is that the 
distribution of the mother boundary is universal, and that the baby 
boundaries are non-universal.
We should thus notice the mother part of the distribution.
Fig.\ref{Fig:BVD_Mother_Critical} shows the volume of the mother boundary 
versus the geodesic distance at the critical points $\kappa_{2}=1.2581$ and 
$1.2770$ with $N_{4}=32K$ and $64K$, respectively.
In the case of $N_{4}=32K$, we estimate the scaling parameter 
$\alpha = 2.4(1)$\footnote{A number inside of $()$ is a systematic error 
from choices of the range of $D$. 
The determination of the range in which the mother volume expects a 
power law behavior of $D$ is more or less ambiguous.} for the range 
$7 \leq D \leq 10$.
In the case of $N_{4}=64K$ we also estimate the scaling parameter 
$\alpha = 3.0(1)$ for the range $7 \leq D \leq 12$.
From Fig.\ref{Fig:BVD_Critical} we can also roughly estimate the scaling 
variable of the baby boundaries, $x=V/D^{4.0}$.
We cannot measure the scaling parameter for the baby boundaries with the 
same accuracy for the mother boundary.

In order to discuss the universality of the scaling relations, we assume 
the distribution function in terms of a scaling parameter, $x=V/D^{\alpha}$, 
as 
\begin{equation}
\rho(x,D) = \frac{1}{D^{\alpha}} x^{a_{1}}e^{-a_{2}x},
\end{equation}
where $a_{1}$ and $a_{2}$ are some constants.
We obtain the following expectation value of the boundary three-dimensional 
volume appearing at distance $D$: 
\begin{equation}
<V^{(3)}(D)> = \frac{1}{N} \int_{v_{0}}^{\infty} dV \; V \; \rho(x,D),
\label{eq:expectationV}
\end{equation}
where $v_{0}$ denotes the UV cut-off of the boundary three-dimensional 
volume and $N = \int_{v_{0}}^{\infty} dV \; \rho(x,D)$ is a normalization 
factor.

If $a_{1} > -1$, the right-hand side of eq.(\ref{eq:expectationV}) converges 
with $v_{0} \rightarrow 0$, as follows: 
\begin{equation}
<V^{(3)}(D)> = \frac{\Gamma(a_{1} + 2)}{a_{2} \Gamma(a_{1} + 1)} D^{\alpha},
\end{equation}
and gives a finite fractal dimension, $d_{f} = \alpha + 1$.
If $a_{1} \leq -1$, the right-hand side of eq.(\ref{eq:expectationV}) depends 
on the cut-off $v_{0}$, and cannot give a finite value when 
$v_{0} \rightarrow 0$.
In this region, $a_{1} \leq -1$, we can estimate the dominate part of this 
integration near to $v_{0} \rightarrow 0$ for four regions, as follows: 
\begin{equation}
<V^{(3)}(D)> \;\;\; \sim \;\;\; \left\{
 \begin{array}{ll}
  \left\{ a_{2} \log{ \left( \frac{D^{\alpha}}{v_{0}} \right) } \right\}^{-1} 
  \;\; D^{\alpha} & 
  \mbox{for} \;\;\; a_{1} = -1 \\[0.5cm]
  v_{0}^{ -(a_{1} + 1)} \;\;\; D^{\alpha (a_{1} + 2)} & 
  \mbox{for} \;\;\; -2 < a_{1} < -1 \\[0.5cm]
  v_{0} \log{ \left( \frac{D^{\alpha}}{v_{0}} \right) } D^{-\alpha} &
  \mbox{for}\;\;\; a_{1} = -2 \\[0.5cm]
  v_{0} D^{-\alpha} &
  \mbox{for}\;\;\; a_{1} < -2.
 \end{array} \right.
\end{equation}
It is clear that $<V^{(3)}(D)>$ goes to zero when $v_{0} \rightarrow 0$ 
whenever $a_{1} \leq -1$.
That is to say, the distribution with $a_{1} \leq -1$ is non-universal.

We also give fractal dimensions by eq.(\ref{eq:fracta_dimension}), 
\begin{equation}
d_{f} = \alpha + 1 \;\;\; 
\mbox{for}\;\;\; -2 < a_{1},
\end{equation}
\begin{equation}
d_{f} = \alpha + 1 \;\;\; \mbox{(with logarithmic correction)} \;\;\;
\mbox{for} \;\;\; a_{1} = -2,
\end{equation}
and 
\begin{equation}
d_{f} = 1 - \alpha (a_{1} + 1) \;\;\; 
\mbox{for}\;\;\; a_{1} < -2.
\end{equation}

We can extract the function of $\rho(x,D)$ from Fig.\ref{Fig:BVD_Critical}, 
and actually find $a_{1} \simeq 0.5$ for the mother volume.
As a result, the mother volume distribution is universal.
On the other hand, we find $a_{1} \simeq -2.0$ for the baby volumes.
It is entirely fair to say that the distribution of the baby volumes is 
non-universal because it depends on the lattice cut-off ($v_{0}$).
A similar non-universal distribution of the boundary volume also appears in 
the weak-coupling phase.

\subsection{The weak-coupling phase}
\begin{figure}
\centerline{\psfig{file=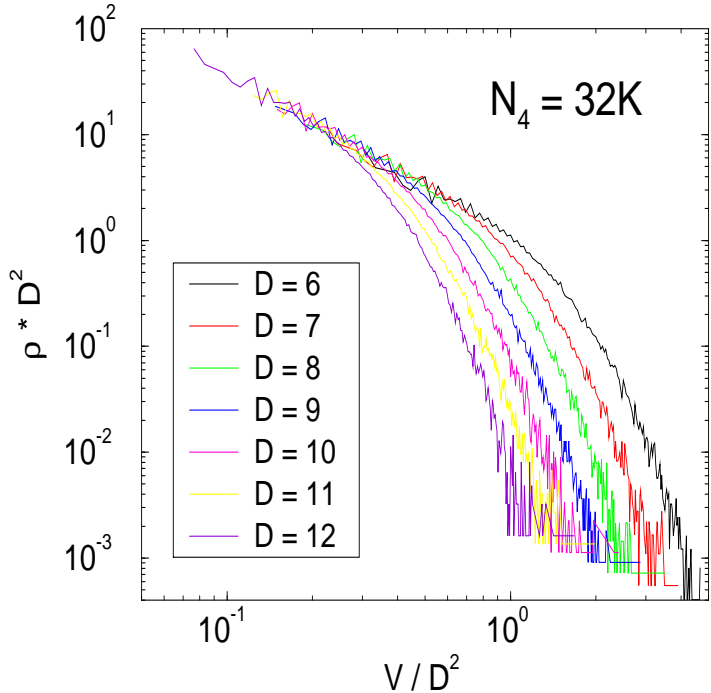,height=10cm,width=10cm}}
\caption
{
Boundary volume distribution in the weak-coupling phase, $\kappa_{2}=2.0$ 
with $N_{4} =32K$ with double-log scales.
}
\label{Fig:BVD_Weak_32K}
\end{figure}

In the weak-coupling phase (we chose $\kappa_{2}=2.0$), as mentioned before, 
the dynamically triangulated manifold resembles an elongated branched 
polymer; thus, we cannot observe the mother universe at all.
Fig.\ref{Fig:BVD_Weak_32K} shows the boundary volume distributions with 
$x = V/D$ as a scaling variable. 
%
%
Moreover, from this figure we can safely state that the 
$\rho(x)$ scales as 
\begin{equation}
\rho (x) \times D \propto x^{-2.0} e^{-x},
\end{equation}
and that the distribution is non-universal.

For all measurements of the boundary volume distributions we prepared 
about 100 independent configurations on the average and 100 four-simplexes 
randomly in each configuration as the origin of the geodesic distance.
%

\section{Summary and discussions}
We investigated four-dimensional spacetime structures generated by dynamical 
triangulation using the concept of the geodesic distance.
In an analogy of the loop length distribution in two-dimensional case, 
the scaling relations in four dimensions are discussed for the three phases 
$i.e.$ the strong-coupling phase, the critical point and the weak-coupling 
phase.
The boundary volume distributions ($\rho(x,D)$) give some basic scaling 
properties on the ensemble of Euclidean space-times described by the 
partition function eq.(\ref{eq:PF}).
In the case of two dimensions we know that the loop length distribution 
shows excellent agreement with an analytical prediction,\cite{KKMW,TY} 
and that both of the baby loops and the mother loop show scalings with the 
same parameter ($x=L/D^{2}$).
However, since the loop length distribution of the baby boundary depends on 
the lattice cut-off, we can recognize that it is non-universal. 

The question is the boundary volume distributions in the four dimensions.
In the strong-coupling limit ($\kappa_{2}=0$) we find that the mother 
part of the boundary volume distribution ($\rho(x,D)$) scales with 
$x=V/D^{d_{f}-1}$ as a scaling variable.
In this phase there is general agreement that the four-dimensional 
dynamically triangulated manifold seems to be a $d_{f}$-sphere($S^{d_{f}}$).
What is important for the constructive definition of 4D simplicial quantum 
gravity is the fact that the boundary volume distribution of the mother 
universe in the strong phase gradually changes into that at the critical 
point.
The fluctuations of the spacetime growth with 
$\kappa_{2} \to \kappa_{2}^{c}$ and at the same time the distribution of the 
baby boundaries comes to scaling (see Figs.\ref{Fig:BVD_Strong} and 
\ref{Fig:BVD_Critical}).
%

At the critical point in the case of four dimensions we have obtained 
similar boundary volume distributions to the two-dimensional case.
However, two different scaling relations are found: one is for the mother 
boundary ($x=V/D^{2.4}$ and $a_{1} \sim 0.5$) and the other is for the baby 
boundaries ($x=V/D^{4.0}$ and $a_{1} \sim -2$).
For the reasons mentioned above, the boundary volume distribution of the 
mother universe seems to be universal (i.e. there is no dependence on the 
lattice cut-off) and that of the baby universes seems to be non-universal 
(i.e. there is a dependence on the lattice cut-off).
We should notice that there exists a considerable size dependence of the 
scaling variable.
In the case of $N_{4}=32K$ and $64K$ we obtain 
$x=V/D^{2.4(1)}$ and $V/D^{3.0(1)}$, respectively.
As of now, these are all of the data in terms of the scaling variable.
There is still room for further investigation.
%

In the weak-coupling phase we obtained elongated manifolds, 
in other words, the branched polymers.
In this phase no mother universe exists and the boundary volume 
distribution of the baby universes shows a scaling relation with scaling 
parameters $\alpha \simeq 1.0$ and $a_{1} \simeq 2.0$ 
(see Fig.\ref{Fig:BVD_Weak_32K}).
However, we recognize that the distribution of the baby boundaries is 
non-universal.
%

We think that the cut-off dependence of the distribution of the baby 
boundaries in all the ranges of $\kappa_{2}$ is reflected in the existence 
of the Planck scale in this model.
In ref.13), we see that the effective curvature increases steeply (diverges) 
when $D \rightarrow 0$ in all phases of $\kappa_{2}$: 
$0.8$(strong phase) $\leq \kappa_{2} \leq 1.5$ (weak phase) and the authors 
suggest the existence of a 'plankian regime'; $D \sim 7.5$ on grounds of the 
divergence. 
Our data of the baby boundaries are consistent with their view; also, 
the data of the mother and baby boundaries clear the origin of the 
scaling relations, which has been argued in ref.13),14).
It seems reasonable to suppose that the lattice model described by 
eq.(\ref{eq:PF}) corresponds to an effective theory which shows a rapid phase 
transition as well as three different types of scaling relations in each 
phase.\cite{BS_4DC,BBKP_4DQG}
In ref.16) we can find a discussion of another model which may have a 
continuous transition.
For the present, it remains an unsettled question as to how to construct a 
consistent lattice model including ``real'' four-dimensional gravity in 
the continuum limit.

However, our numerical results are expected to be a first step to research 
the universal scaling relations in two-, three- and four-dimensional 
simplicial quantum gravity. 
%

\section{Acknowledgments}
We would like to thank H.Kawai, N.Ishibashi, S.R.Das, J.Nishimura, 
M.Inaba and H.Hagura for fruitful discussions. 
We are also grateful to the members of the KEK theory group and N.Matsuhashi 
for advice and hospitality. 
Numerical calculations were performed using the HP 700 series at 
KEK and the Convex SPP at Tokyo Institute of Technology. 
Some of the authors (T.H., T.I. and N.T.) were supported by Research 
Fellowships of the Japan Society for the Promotion of Science for Young 
Scientists.

\end{document}